\documentclass[useAMS,usenatbib]{mn2e}
\usepackage{graphicx}

\def\l{~$\lambda$}

\def\hea{He\,{\sc i}}
\def\heb{He\,{\sc ii}}
\def\nc{N\,{\sc iii}}
\def\nd{N\,{\sc iv}}
\def\ne{N\,{\sc v}}

\def\sid{Si\,{\sc iv}}

\def\kms{km\,s$^{-1}$}

\def\lsol{L$_{\odot}$}
\def\msun{M$_{\odot}$}

\def\snr{$S/N$}

\def\xsho{{X-shooter}}

\def\gtrsim{\mathrel{\hbox{\rlap{\hbox{\lower3pt\hbox{$\sim$}}}\hbox{\raise2pt\hbox{$>$}}}}}
\def\lesssim{\mathrel{\hbox{\rlap{\hbox{\lower3pt\hbox{$\sim$}}}\hbox{\raise2pt\hbox{$<$}}}}}

\title[R144 revealed as a double-lined spectroscopic binary]{R144 revealed as a double-lined spectroscopic binary\thanks{Based on observations collected at the European Southern Observatory (ESO) under GTO program IDs 085.D-0704 and 086.D-0446.}}
\author[Sana et al. ]{H. Sana,$^{1}$\thanks{E-mail: H.Sana@uva.nl}
T.~van Boeckel,$^{1}$
F.~Tramper,$^{1}$
L.E.~Ellerbroek,$^{1}$
A.~de Koter,$^{1,2,3}$\newauthor
L.~Kaper,$^{1}$
A.F.J.~Moffat,$^{4}$
O.~Schnurr,$^{5}$
F.R.N.~Schneider,$^{6}$
D.R.~Gies$^{7}$\\
$^{1}$    Astronomical Institute Anton Pannekoek, 
          Amsterdam University,  
          Science Park 904, 1098~XH, 
          Amsterdam, The Netherlands\\
$^{2}$    Utrecht University,
          Princetonplein 5, 3584CC,
          Utrecht, The Netherlands\\
$^{3}$    Instituut voor Sterrenkunde, 
         Universiteit Leuven, 
         Celestijnenlaan 200 D, 
         3001, Leuven, Belgium\\
$^{4}$    D\'epartment de Physique, 
          Universit\'e de Montr\'eal and
          Centre de Recherche en Astrophysique du Qu\'ebec,
          C. P. 6128, succ. centre-ville, 
          Montr\'eal (Qc) H3C 3J7, Canada\\
$^{5}$    Leibniz Institut f\"ur Astrophysik Potsdam (AIP), 
          An der Sternwarte 16, 
          14482 Potsdam, Germany \\
$^{6}$    Argelander-Institut f\"ur Astronomie, 
          Universit\"at Bonn, 
          Auf dem H\"ugel 71, 
          53121 Bonn, Germany\\
$^{7}$    Center for High Angular Resolution Astronomy and 
          Department of Physics and Astronomy, 
          Georgia State University,\\ 
          P.O. Box 4106, Atlanta, 
          GA 30302-4106, USA}
\begin{document}

\date{Accepted 1988 December 15. Received 1988 December 14; in original form 1988 October 11}

\pagerange{\pageref{firstpage}--\pageref{lastpage}} \pubyear{2002}

\maketitle

\label{firstpage}

\begin{abstract}
R144 is a WN6h star in the 30 Doradus region. It is suspected to be a binary because of its high luminosity and its strong X-ray flux, but no periodicity could be established so far. Here, we present new \xsho\ multi-epoch spectroscopy of R144 obtained at the ESO Very Large Telescope (VLT). We detect variability in position and/or shape of all the spectral lines.  We measure radial velocity variations with an amplitude larger than 250~\kms\ in \nd\ and \ne\ lines. Furthermore, the \nc\ and \ne\ line Doppler shifts are anti-correlated and the \nd\ lines show a double-peaked profile on six of our seven epochs. We thus conclude that R144 is a double-lined spectroscopic binary. Possible orbital periods range from 2 to 6 months, although a period up to one year is allowed if the orbit is highly eccentric. We estimate the spectral types of the components to be WN5-6h and WN6-7h, respectively. The high luminosity of the system ($\log L_\mathrm{bol}/L_\odot \approx 6.8$ ) suggests a present-day total mass content in the range of about 200 to 300~\msun, depending on the evolutionary stage of the components. This makes R144 the most massive binary identified so far, with a total mass content at birth possibly as large as 400~\msun.  We briefly discuss the presence of such a massive object 60~pc away from the R136 cluster core in the context of star formation and stellar dynamics. 
\end{abstract}


\begin{figure*}
  \includegraphics[width=16cm]{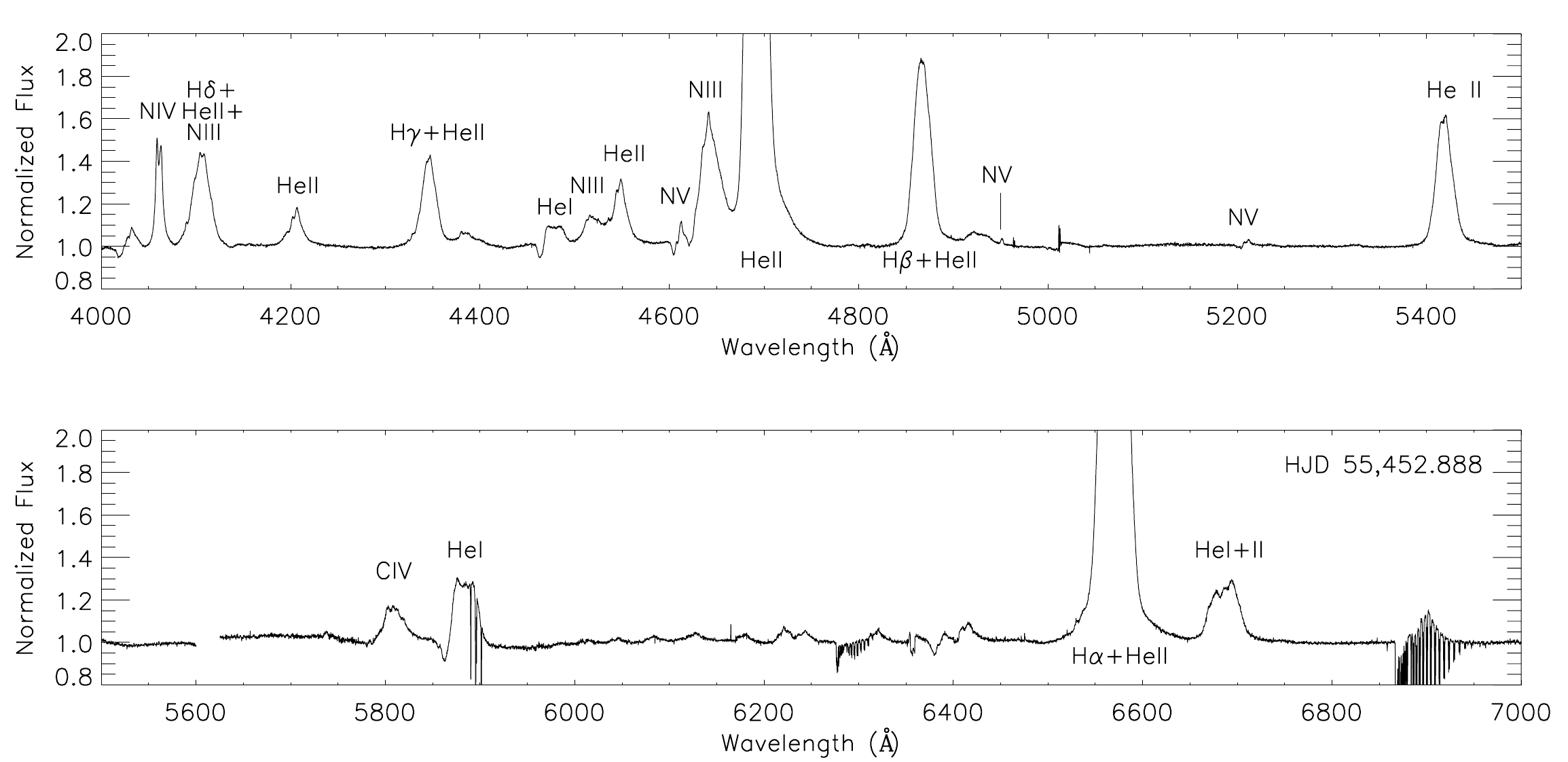}
 \caption{Optical spectrum of R144 at the epoch of largest RV separation (\#2 in Table~\ref{tab: journal}). The main spectral lines are identified. }
 \label{fig: spec}
\end{figure*}

\section{Introduction}

Massive stars strongly contribute to the light from distant star-forming regions and dominate the production of ionizing radiation. The most massive stars  are important contributors to  the feedback from stars on the interstellar medium \citep{Cro07}. Yet, how massive a star can be is still poorly constrained.  In the quest for a firm grip on the maximum stellar mass limit, hydrogen-rich Wolf-Rayet (WNh) stars have been identified as the most massive stellar objects \citep{dKHH98}, with typical dynamical masses over 75~\msun\ \citep{RdBN04_mnras, GGM06_mnras, NGB08, SCC08, SMVF09}. \citet{CSH10} have recently re-analysed known WNh stars in R136, the central cluster in 30~Dor, and derived probable present-day masses up to 250~\msun. Other very massive WNh stars are found in apparent isolation in the 30 Dor region, among them R145 \citep{SMVF09} and VFTS~682 \citep{BVG11_mnras}. Unless these stars are all ejected from the cluster core through dynamical interactions or supernova explosions, their presence challenges the view in which massive stars formed along with a cortege of lower mass stars in a dense cluster environment.

With $v=11.15$ \citep{bat99}, the WN6h star R144 (RMC 144, Brey 89, BAT99-118, HD 38282) in  30 Dor is the visually brightest Wolf-Rayet star in the Large Magellanic Cloud (LMC) and it is one of the early-type objects isolated from the R136 core. Because of its brightness and its strong X-ray emission, R144 has been proposed as a  binary candidate, but no definitive proof of the orbital motion has been obtained so far. \citet{Mof89} performed the first multi-epoch spectroscopic campaign on R144 and, in the absence of variability, concluded that the object was likely single. Using a more extended observational data set, \citeauthor{SMSL08} (\citeyear{SMSL08}, \citeyear{SMVF09}) detected  variability in the \heb\l4686 line as well as in broadband polarimetry, although no periodicity could be established. In this work, we report on the detection of R144 as a double-lined spectroscopic binary and  we suggest that the two components are WNh stars with a present-day total mass of up to 300~\msun\ for the binary system. 

\section{Observations and data reduction}\label{sect: obs}

Intermediate-resolution spectra of R144 were collected from April 2010 to April 2011 using the  \xsho\ spectrograph \citep{VDDO11_mnras} mounted on VLT/UT2 at ESO's Paranal observatory. In total R144 was observed at seven epochs, allowing us to explore time scales of days, weeks and months.  \xsho\ provides us with complete optical and near-infrared (NIR) wavelength coverage from the $U$ to the $K$-band by combining three instrument arms, each operating as a separate spectrograph. Given the brightness of R144 and our scientific aims, the narrowest slit of each arm was used, providing a resolving power of 9100, 17400 and 11000 in the UVB, VIS and NIR arms, respectively. The observations were carried out in nodding mode with a nod throw of 5\arcsec\ along the parallactic angle. Exposures of 90s (resp. 60s) for the UVB and VIS arms (resp. NIR arm) were obtained at each nodding position. Depending on the weather conditions, two to four nodding cycles were performed, yielding a typical signal-to-noise ratio per resolution element ($S/N$) of over 200 in the UVB and VIS arms and over 100 in the NIR arm. The journal of the observations is given in Table~\ref{tab: journal}.

The data were reduced and flux calibrated using the \xsho\ pipeline running under the Reflex workflow environment \citep{MGR10_mnras}. The spectra of each arm were then normalised to the continuum by fitting a polynomial to continuum regions. An example of the obtained spectra is presented in Fig.~\ref{fig: spec}. The R144 spectrum is dominated by strong emission lines of hydrogen, helium and nitrogen that are characteristic of WNh stars \citep{SSM96}. \hea\ lines show clear P-Cygni profiles. We also  detect \ne\ P-Cygni lines at 4604 and 4620~\AA\ as well as a single-lined \ne\l4944 emission.

\begin{table}
 \centering
 \begin{minipage}{75mm}
  \caption{Journal of the observations. Cols. 3 to 5 give the primary ($v_1$) and secondary ($v_2$) radial velocities measured from  \ne\l4944 ($v_1$ only) and \nd\l4058 ($v_1$ and $v_2$).}
  \label{tab: journal}
  \begin{tabular}{@{}ccccc@{}}
  \hline
  \#&  HJD$^a-$  & \ne\l4944    & \multicolumn{2}{c}{\nd\l4058}\\
    & 2,400,000  & $v_1$ (\kms) & $v_1$ (\kms)  & $v_2$ (\kms) \\
\hline
1& 55,309.478 &  357  &   336  &   56 \\
2& 55,452.888 &  419  &   380  &   65 \\
3& 55,580.589 &  151  &    64  &  319 \\
4& 55,585.650 &  137  &    57  &  319 \\
5& 55,589.681 &  159  &    61  &  308 \\
6& 55,604.525 &  387  &   360  &   59 \\
7& 55,673.479 &  360  &   342  &   73 \\
 \hline
\end{tabular}\\
{\sc notes -} a. Heliocentric Julian date at the beginning of the exposure.\\
\end{minipage}
\end{table}

\begin{figure}
  \includegraphics{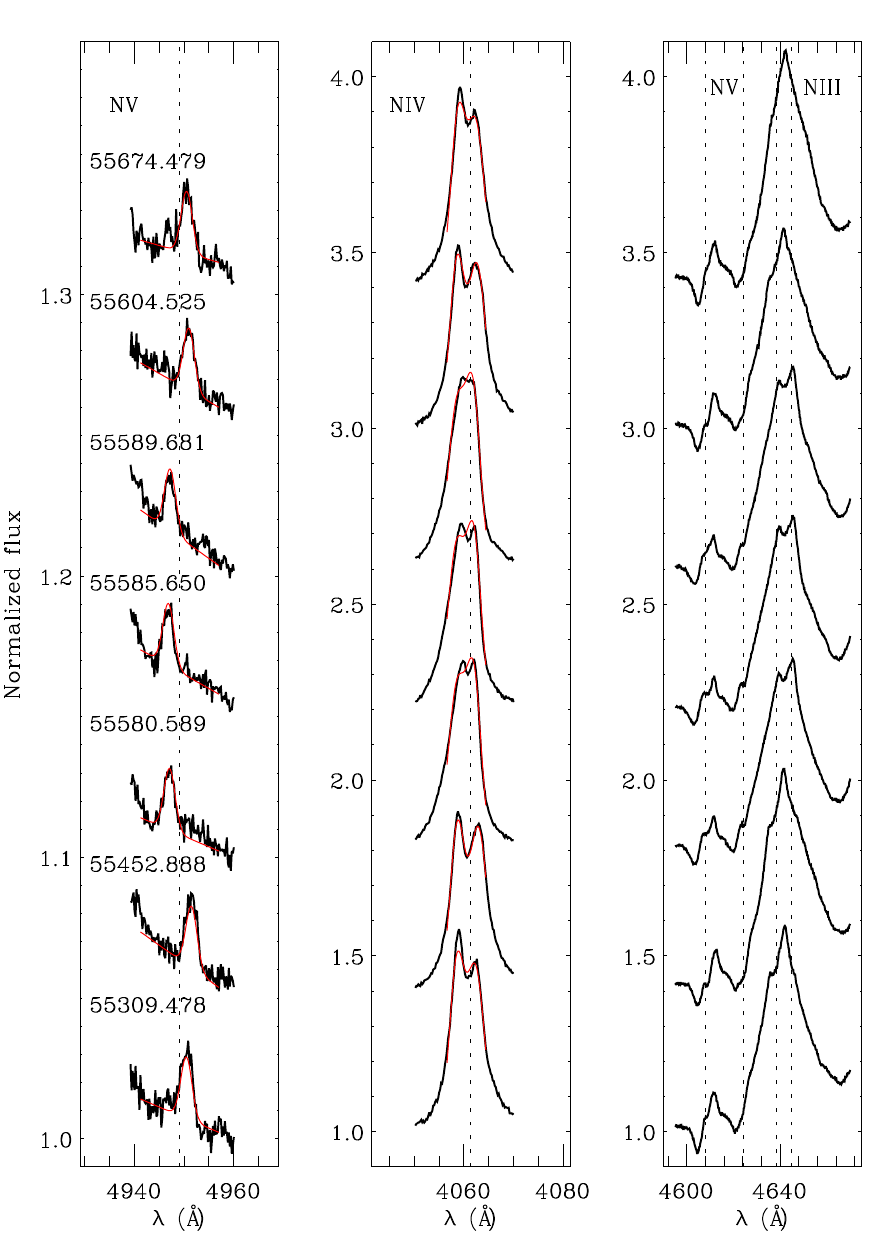}
 \caption{\ne\l4944, \nd\l4058 and \nc\l4640-42 lines at different epochs. The vertical dashed lines indicate the rest wavelength  of \ne, \nd\ and of both \nc\ components, shifted to the LMC rest-frame (i.e., shifted by $+270$~\kms). Red lines show our best-fit RV measurements from Table~\ref{tab: journal}.}
 \label{fig: spec_rv}
\end{figure}

\section{Variability and RV measurements}\label{sect: var}

We search for line profile variability in the time series formed by our spectra by visual inspection and by using a temporal variance spectral analysis \citep[TVS,][]{FGB96}. Significant line profile variability is detected in most of the spectral lines. The associated TVS spectrum is often double-peaked, which is indicative of large radial velocity (RV) variations, hence of a potential binary motion. We will return to this aspect later. 

Fig.~\ref{fig: spec_rv} shows a selection of \nc, \nd\ and \ne\ lines observed during our campaign.  \ne\l4944 has a single component whose position varies by over 4~\AA. \nd\l4058 has single comparable components from each of the two stars, hence the line is double-peaked with the two peaks changing in position. \nc\l4640-42 effectively has two components from each star, thus a total of four. The line profile changes from a single peak to double as a result of line blending. 

As illustrated in Fig.~\ref{fig: spec_rv}, the \ne\ line and the dominant \nc\ components present anti-correlated RV variations. Further visual inspection of our spectra indicates that the positions of the \sid\ and \ne\ lines also vary in opposite direction. 
Finally, the widths and centroids of hydrogen and helium lines are found to be variable, but the lines remain single-peaked, probably due to their intrinsically broad nature. 

\begin{figure}
  \includegraphics[width=8cm]{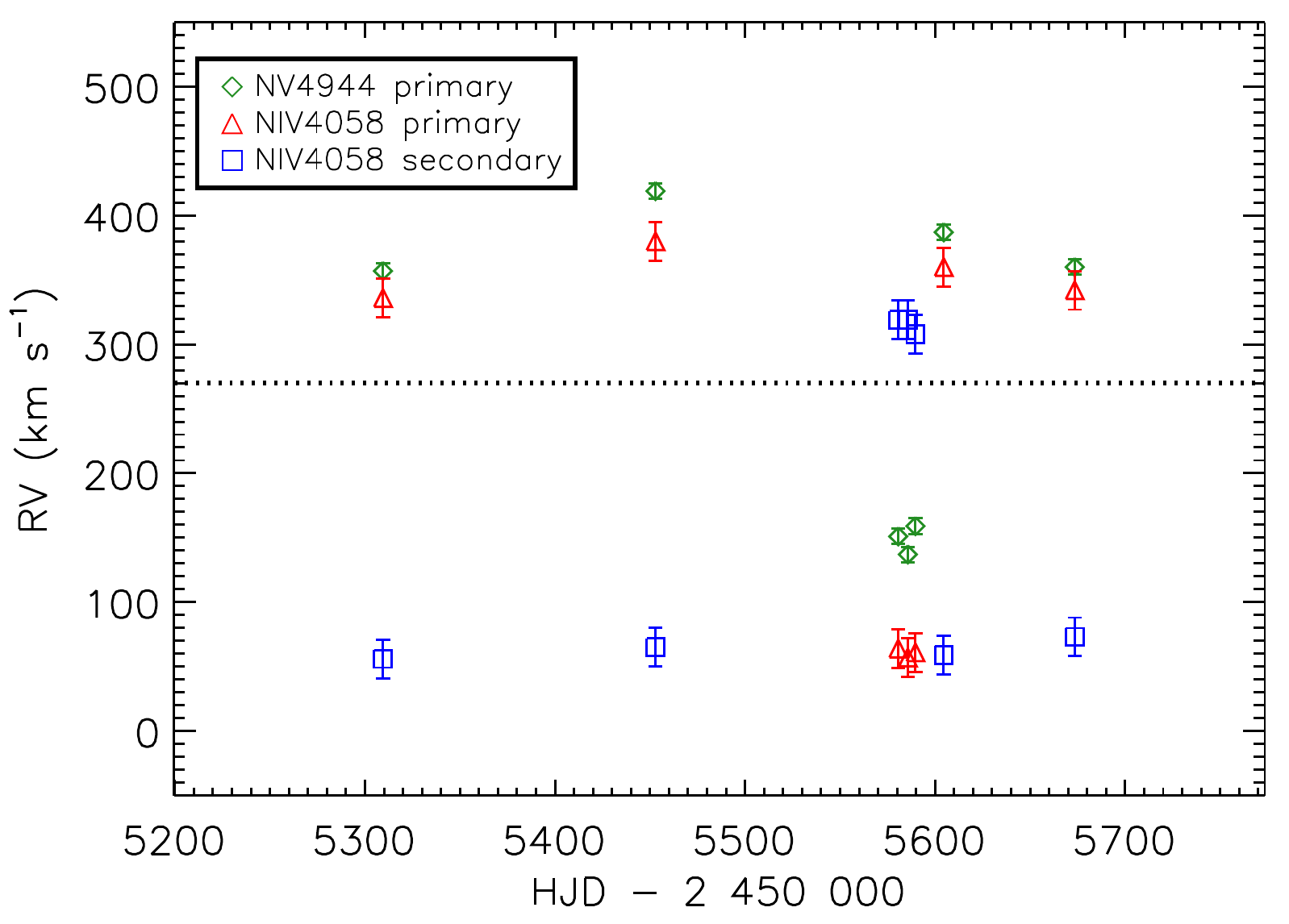}
 \caption{\ne\l4944 and \nd\l4058 RV measurements. The dotted line indicates the average velocity of stars in 30 Dor.}
 \label{fig: rv}
\end{figure}

To quantify the amplitude of the RV variations, we use the fitting method described in \citet{SdKdM13_mnras}, which adjusts Gaussian(s) to line profiles accounting simultaneously for all the observations.  We separately adjust the \ne\l4944 and \nd\l4058 line profiles and we weight the fit by the \snr. The best fit Gaussian(s) are overlaid in Fig.~\ref{fig: spec_rv}. The associated RV measurements are listed in Table~\ref{tab: journal} and displayed in Fig.~\ref{fig: rv}. A local renormalisation has been applied for \ne\l4944 given that the line is positioned in the wing of \hea\l4922. 

The  \ne\l4944 line is well fitted by a single Gaussian and displays a peak-to-peak RV variation of  280~\kms. The typical measurement errors are about 6~\kms.  In the following, we arbitrarily refer to the component associated with the \ne\l4944 emission as the primary, hotter star in the system.

The wings of the  \nd\l4058  line are not properly represented by a Gaussian profile and we limit the fit to the central part of the line only. Two Gaussians are used and Fig.~\ref{fig: spec_rv} only shows the sum of the fitted Gaussians. The  \nd\l4058 RVs somewhat depend on the wavelength range adopted for the fit. Combined with the fact that this line profile is not Gaussian, we estimate error bars of the order of 15-20~\kms\ on the RV values. While the peak of the blue component of \nd\l4058 remains more or less at the same location, its width and the shape of its blue wing changes. The best fitting solution indicates thus  that the two components of \nd\l4058\ do actually exchange position at epochs \#3 to 5 with respect to the other epochs. In this case, 
the primary and secondary components display a peak-to-peak RV variation of about 320 and 260~\kms, respectively.

Overall there is a reasonable agreement between the primary RVs obtained from the two lines. \nd\l4058 seems to display a slightly larger RV amplitude than \ne\l4944 although the obtained values remain within the uncertainties of the RV measurement method. 

\section{The nature of R144}\label{sect: r144}

\subsection{Spectral type }
We first classify the composite spectrum of R144 as if it were a single star. The criteria of \citet{SSM96}, which are based on ratios of peak intensities of H, He, C and N lines, unambiguously confirm the WN6h type of the composite spectrum. Given the blue-shifted weak absorptions in each of the \hea\ and \heb\ emission lines, the final composite spectral type is  WN6ha.

As mentioned earlier, the dominant component of the \nc\l4640-42 lines moves opposite to \ne\l4944 (Fig.~\ref{fig: spec_rv}). The fainter \nc\l4510-14-18 multiplet shows a similar behaviour. This suggests that the \nc\ lines are predominantly associated with the secondary, cooler star. In order to qualitatively reproduce the profile variations of the \nc\l4640-42 complex, the primary, hotter star is still expected to make a small contribution to \nc\l4640-42, possibly with a peak intensity no larger than half that of the secondary. A peak intensity of a third of that of the secondary, or less, would imply a WN5h classification, which we consider to be within uncertainties. A complete spectral disentangling with a larger data set is, however, needed to properly quantify the amount of \nc\l4640 in the primary star. 

If most of the \nc\ emission in the R144 spectrum originates in the secondary star, accounting for the dilution of the secondary \nc\ line by the primary light suggests that the secondary may be of a slightly later spectral type. We thus adopt WN5-6h + WN6-7h as our best estimate of the system  spectral types.

\subsection{Orbital constraints}

\subsubsection*{Mass ratio}

The mass ratio of a double-lined binary system can in principle be constrained directly from the RV measurements of the two components.  We use the orthogonal least-square regression method of \citet{SGR06_219} to fit the $v_1$ vs.\ $v_2$ relation, allowing for different apparent systemic velocities for the two components. We obtain a mass ratio of $M_2/M_1=1.17 \pm 0.06$, which implies that the primary, hotter star is the less massive star in the system. While it is possible that systematics in the RV measurements of the \nd\l4058 line might actually invert the mass ratio, our results clearly indicate that the masses of the two components do not differ by more than a few tens of percent from one another.

\subsubsection*{Orbital period} \label{sect: period}
Our data only cover one epoch of large RV variations and do not allow us to constrain the periodicity of the system. The large RV change ($>200$~\kms) between observations 5 and 6, separated by only two weeks, suggests either a relatively short period or a significant eccentricity, possibly both. Combining the epoch of variability observed in our \xsho\ data with those in \citeauthor{SMSL08} (\citeyear{SMSL08}, \citeyear{SMVF09})  does allow us to obtain a first constraint on the orbital period. Significant variability is observed by \citeauthor{SMSL08} at $t_0 \approx 2,452,630$ and at $t_1 \approx 2,453,000$, hence a $\Delta t$ of about 370~d. Our  RV variations lie at $t \approx 2,455,586$ (Table~\ref{tab: journal}), hence almost exactly $8 \times \Delta t$ from epochs $t_0$ (resp.  $7 \times \Delta t$ from $t_1$). This strongly suggests that the  maximum orbital period of R144 is about 370~d.

Submultiple integer values (i.e., $P\sim 370/n$ with $n=1, 2, 3,\dots$) constitute a family of  possible solutions, where $n=4, 5$ ($P \sim 90$ or 70 d) and  $n\ge7$ ($P \le 50$~d) are unlikely,  as large RV variations are predicted at epochs where \citet{Mof89}, \citeauthor{SMSL08} and/or our \xsho\ observations show a constant signal. This deduction assumes that the peak of the RV variations is concentrated over a time span of 15 days but the same constraints are obtained by assuming 10 to 20 days.  The most favoured periods are thus about 2, 4, 6 and 12 months. 

The semi-amplitudes of the observed RV variations (Table~\ref{tab: journal}) provide a lower limit on the true semi-amplitudes of the RV curves. For given stellar masses and eccentricity, they constrain the maximum orbital period. Adopting masses of 120 and 140~\msun\ for both components (see Sect.~\ref{sect: masses}),  and eccentricities of 0.3, 0.5 and 0.7, the upper limit on the period is 4.5, 6 and 10.5 months, which matches  the range of possible periods discussed above. 

\section{Discussion}\label{sect: discuss}

\subsection{Stellar mass content}\label{sect: masses}

Mass estimates can be obtained from the spectral types. WN5h stars have typical (present-day) masses in the range 130 to 250~\msun, while  WN6h and WN7h stars have masses between 80 and 130~\msun\ \citep[][and references therein]{CSH10}. This suggests that R144 is formed by two $\gtrsim$ 80~\msun\ stars, making it one of the most massive binary systems known.

The total mass content of R144 can also be estimated from its bolometric luminosity ($L_\mathrm{bol}$). We use the narrow-band photometry of \citet[][$v=11.15$, $b-v=-0.11$]{bat99}. 
We estimate the visual extinction in two different ways: (i) we deredden the \xsho\ flux-calibrated spectrum to the Rayleigh-Jeans slope, yielding $A'_v=1.1 \pm 0.2$; (ii) we use the color excess $E(b-v) \approx 0.21$ and $R'_v=4.12$ \citep[corresponding to $R_V=3.1$,][]{Tur82}, resulting in $A'_v \approx 0.9$. Both extinction estimates agree within the uncertainties. To avoid overestimating the bolometric luminosity, we conservatively adopt the latter value and obtain an absolute visual magnitude $M_v=-8.2$ for the system. The visual magnitude of R144 is thus about 1~mag brighter than other WN6h binaries such as R145 \citep[$M_v \sim -7.2;$][]{SMVF09} and WR20a \citep[$M_V = -7.04;$][]{RCdB05_mnras}. Given that both R145 and WR20a have a total mass content of 160~\msun\ or larger, the brightness of R144 suggests an even larger mass content.

While the bolometric correction remains uncertain without a detailed atmospheric modelling, it is likely in the range 4.0 (a typical early O star) to 4.7 mag \citep[see e.g.][]{BVG11_mnras}, yielding  $\log L_\mathrm{bol}/L_\odot \approx 6.8$ to 7.0. The total luminosity of R144 is thus very similar to that of the other WNh stars in the core of R136 and brighter than VFTS682 in the surrounding nebula. 

Adopting $\log L_\mathrm{bol}/L_\odot = 6.8$ and a mass-luminosity relation appropriate for very massive stars (K\"{o}hler et al., in prep.), we estimate masses of 170 + 205~\msun\ if the stars are on the zero-age main sequence (MS) and of 80 + 95~\msun\ if the stars are close to the MS turn-off, i.e.\ the TAMS.

\subsection{``Isolated'' formation or dynamical ejection}

R144 is located at a projected distance of about 60 pc from the central cluster R136. The presence of very massive stars  outside cluster cores is not predicted by most massive star formation theories, which expect that massive stars form as part of a cluster \citep{ZiY07}. R144 may thus have been ejected from the cluster core, in which case R144 would have needed a (projected) runaway velocity of $\sim$~60~\kms\ to travel from  R136 to its current location in $\approx 10^6$ years.  Because the two components of R144 are both very massive stars (i.e., with a short evolutionary time scale), the supernova kick scenario is unlikely. In the runaway scenario, R144 then needs to have been ejected from R136 through dynamical interactions.

In dense stellar environments such as R136, the most massive stars tend to sink quickly to the cluster center where they dynamically interact  with each other. During such interactions, the less massive object is typically ejected \citep{Heg75} and the standard theory thus requires an even more massive object to have caused the ejection of R144. Such an object is not seen. In principle, this object might already have terminated its evolution despite the young age of the region. However, stars of hundreds of solar masses are all expected to have a similar life-time of about 2.3~Myr (K\"{o}hler et al., in prep.). As other very massive stars are still present in the core of R136, this scenario would require significant fine tuning.

An alternative scenario for ejecting R144 from R136 that does not require such an extremely massive object is offered by recent theoretical work. \citet{FPZ11} calculated that the gravothermal collapse of a cluster core produces, in the core, a binary that is formed by the most massive stars in the cluster. This binary then dominates the dynamical interaction in the core, frequently ejecting other stars. It hardens by each dynamical encounter until it gets ejected as well. In that scenario, R144 is a candidate ``bully binary'', possibly formed in the  gravothermal collapse of the cluster core. This hypothesis is not without drawbacks. First, single stars more massive than the R144 components likely exist in the cluster core \citep{CSH10}, so that the bully binary is not formed by the most massive stars. Second, its orbital period ($\lesssim 370$~d) seems too short compared to the typical periods of systems predicted to form during the collapse \citep[$P \sim 1000$~d,][]{FPZ11}. 

If R144 is not a runaway system, it has thus formed in situ and in relative isolation from R136. One possibility is that R144 is part of an older association along the line of sight toward the northern part of 30 Dor. R144 is indeed located in the vicinity of the luminous WN stars R146 and R147 (projected distance of about 22pc), and within a giant high-velocity/X-ray shell that is most likely a supernova remnant \citep{FTW60, WBS13}. This scenario, however, supposes the entire dispersion of the parent association to explain the low surface density of O- and B-type stars in the vicinity of R144. 

The last scenario, in which R144 has actually formed in situ and independently of a parent cluster or OB association,  would pose a serious challenge to massive star formation scenarios that rely on a dense stellar environment, such as formation through mergers of lower-mass proto-stellar cores  \citep{BB02} or through competitive accretion \citep{BoB06}. The formation of very massive stars in relative isolation is compatible with a fractal vision of star formation, where a giant molecular cloud fragments in various cores which then form stars stochastically. As discussed in \citet{BBE12_mnras}, discriminating between these  scenarios is not only important for our understanding of high-mass star formation but also to discern how nature samples the initial mass function.

\subsection{Evolutionary stage}

While awaiting an accurate determination of the orbital properties, the fact that the primary, hotter star seems to be less massive than the secondary, cooler component is intriguing. A lower-mass primary component is typically the indication of  past or ongoing binary interaction. Given the relatively large separation of the R144 system, the stars are only expected to fill their Roche lobe towards or after the end of the MS. It is, however, possible that a high eccentricity allows for interaction close to periastron passage. The binary orbit may also have been widened by mass loss through stellar winds or as a result of a past mass-transfer event, so that R144 may have been a tighter binary in the past.

Alternatively, the primary star may have been the more massive component at birth. Given its larger mass, it may have entered the WR-wind mass-loss regime earlier than the secondary, thus shedding mass at such an accelerated rate that it became the less massive component in the system. A qualitative comparison with the K\"ohler et al. (in prep.) evolutionary tracks indicate that a 260+175~\msun\ pair at zero-age MS evolves, after 2~Myr, into a 90+120~\msun\ system  with a total present-day luminosity of $10^{6.9}$\lsol\ and, indeed, with a hotter and less massive primary component.

Characterizing both the orbit and the physical properties of the R144 components may further help to elucidate the evolutionary status of the object, hence to better constrain stellar evolution at very high masses. Gathering high quality spectra with a sufficiently good phase coverage to enable the determination of the full orbital solution and to perform spectral disentangling is thus a high priority goal in gaining a better understanding of R144. A linear polarimetry monitoring campaign and a search for atmospheric eclipses may further help to constrain the inclination of the system \citep[e.g.,][]{MMB98_mnras}, thus to provide a direct measurement of the  masses of the two components of R144.

\section*{Acknowledgments}
We are grateful to K. K\"{o}hler and collaborators for providing mass-luminosity relations of very massive stars prior to publication. We also thank the referee, Nolan Walborn, for his help clarifying the text.

\bibliographystyle{mn2e}
\bibliography{/home/hsana/Desktop/Dropbox/Dropbox/literature}

\bsp

\label{lastpage}

\end{document}